\documentclass[11pt,a4paper]{article}
\usepackage{amsmath, amsthm} 
\usepackage{amsfonts}
\usepackage{amssymb,graphicx,psfrag}
\usepackage[bottom]{footmisc}
\usepackage{hyperref}
\usepackage{mathrsfs}
\usepackage{cite}
\usepackage{setspace}
\usepackage{color}
\usepackage{standalone}

\newcommand{\bea}{\begin{eqnarray}}
\newcommand{\eea}{\end{eqnarray}}
\newcommand{\be}{\begin{equation}}
\newcommand{\ee}{\end{equation}}
\newcommand{\ba}{\begin{array}}
\newcommand{\ea}{\end{array}}

\newcommand{\rd}{\mathcal}

\newcommand{\badat}{\begin{alignedat}}
\newcommand{\eadat}{\end{alignedat}}
\newcommand{\bitm}{\begin{itemize}}
\newcommand{\eitm}{\end{itemize}}
\newcommand{\bmat}{\begin{pmatrix}}
\newcommand{\emat}{\end{pmatrix}}
\newcommand{\bse}{\begin{subequations}}
\newcommand{\ese}{\end{subequations}}

\newcommand{\ep}{\epsilon}

\def\ndelta{\delta\hspace{-0.50em}\slash\hspace{-0.05em} }

\def\nn{\nonumber}

\def\p{\partial}
\def\eps{\epsilon}

\numberwithin{equation}{section}

\definecolor{Geo}{rgb}{0.1,0,0.75}

\def\hybrid{\topmargin -10pt    \oddsidemargin 0pt
        \headheight 0pt \headsep 0pt
       \textwidth 6.25in       
      \textheight 9.5in       
        \marginparwidth .875in
        \parskip 5pt plus 1pt   \jot = 1.5ex}
\hybrid
\numberwithin{equation}{section}
\numberwithin{table}{section}
\begin{document}

\thispagestyle{empty}

\rightline{\small}

\vskip 3cm
\noindent
\begin{spacing}{1.5}
\noindent
{\LARGE \bf  Liouville theory  
beyond the cosmological horizon
 }
 \end{spacing}
\vskip .8cm
\begin{center}
\linethickness{.06cm}
\line(1,0){447}
\end{center}
\vskip .8cm
\noindent
{\large \bf Geoffrey Comp\`ere \\ [.09cm] Laura Donnay  \\ [.09cm] Pierre-Henry Lambert \\ [.09cm] Waldemar Schulgin}

\vskip 0.2cm

\noindent{\em 
Universit\'e Libre de Bruxelles and International  Solvay
Institutes}
\vskip -0.15cm
\noindent{\em
ULB-Campus Plaine CP231}
\vskip -0.15cm
\noindent{\em \hskip -.05cm B-1050 Brussels, Belgium}
\vskip -0.10cm
\noindent{\tt \hskip -.05cm gcompere, ldonnay, pilamber, waldemar.schulgin AT ulb.ac.be }
\vskip1cm

\vskip 1cm

\vskip0.6cm

\noindent {\sc Abstract:} 
The dS/CFT correspondence postulates the existence of a Euclidean CFT dual to a suitable gravity theory with Dirichlet boundary conditions asymptotic to de Sitter spacetime. A semi-classical model of such a correspondence consists of Einstein gravity with positive cosmological constant and without matter which is dual to Euclidean Liouville theory defined at the future conformal boundary. Here we show that Euclidean Liouville theory is also dual to Einstein gravity with Dirichlet boundary conditions on a fixed timelike slice in the static patch. Intriguingly, the spacetime interpretation of Euclidean Liouville time is the physical time of the static observer. As a prerequisite of this correspondence, we show that the asymptotic symmetry algebra which consists of two copies of the Virasoro algebra extends everywhere into the bulk.

\newpage

\tableofcontents

\section{Introduction and outline}

In 1995, some years before the advent of the AdS/CFT correspondence, it has been noticed that three-dimensional Einstein gravity with negative cosmological constant can be rewritten as Lorentzian Liouville theory defined on the conformal  boundary cylinder of AdS, upon imposing suitable Dirichlet-type boundary conditions \cite{Brown:1986nw,Coussaert:1995zp}. The Hamiltonian reduction procedure is achieved in two steps, with the non-chiral WZW model as an intermediate theory. In retrospect, this provided a first toy model of a conformal field theory that is classically equivalent to gravity in AdS, before string proposals \cite{Maldacena:1997re} and higher spin proposals \cite{Gaberdiel:2012uj} were made.

Given the analytic continuation relating anti-de Sitter to de Sitter spacetime, it comes as no surprise that one can similarly rewrite Einstein gravity with positive cosmological constant (with similar Dirichlet-type boundary conditions) in terms of Euclidean Liouville theory \cite{Cacciatori:2001un}. 
More precisely, the Einstein-Hilbert action reduces to two copies of Euclidean Liouville theory, the first defined on the future boundary $\mathcal I^+$ and the second on the past boundary $\mathcal I^-$, since these boundaries border the complete spacetime bulk. However, bulk null geodesics connect any point on the sphere $\mathcal I^-$ to the antipodal point on the sphere $\mathcal I^+$. It has been argued, then, that the formulation of a full-fledged dual quantum theory, a ``dS/CFT correspondence'', would only require one boundary \cite{Strominger:2001pn}. No UV complete string embedding of such a dS/CFT correspondence has been formulated so far but proposals using higher spins have been made \cite{Anninos:2011ui}. 

In the dS/CFT proposal \cite{Strominger:2001pn}, the holographic screen where the CFT would be best defined is the future (or past) conformal boundary. There, one can define the asymptotic symmetries, whose complexification consist of two copies of the Virasoro algebra. One can also define  the conformal dimensions and correlation functions of the  operators dual to bulk fields. The presence of the cosmological horizon of a thermal and entropic nature \cite{Gibbons:1977mu} between the static observer and the conformal boundary however raises questions on whether the holographic description extends all the way to the static observer. In addition, even though one can define the Virasoro central charges to be positive, the semi-classical spectrum of zero modes, which corresponds to spinning conical defects  \cite{Deser:1983nh}, is complex, which challenges the existence of a Hilbert space with a unitarity inner product. Such issues were further discussed in the literature \cite{Witten:2001kn,Bousso:2001mw,Klemm:2001ea,Spradlin:2001pw,Dyson:2002nt,Balasubramanian:2002zh}. Other holographic scenarios were also proposed \cite{Alishahiha:2004md,Freivogel:2006xu,Anninos:2011af}.

In this paper, we first point out that in the case of the three-dimensional Einstein gravity without matter the asymptotic symmetry group is not limited to act at the conformal boundary. Instead, one can extend the notion of the ``asymptotic'' symmetries and the associated conserved charges anywhere into the bulk. This fact holds independently of the sign of the cosmological constant. A convenient way to define the generators everywhere in the bulk makes use of Eddington-Finkelstein type coordinates which were  thoroughly used e.g. in \cite{Barnich:2012aw}. As a result, the conformal group acts naturally in the static patch beyond the cosmological horizon. This provides consistent boundary conditions (which are compatible with conformal symmetry) on any fixed radial slice and in particular close to the horizon.

It is then natural to perform the Hamiltonian reduction of Einstein gravity in the static patch, taking as a boundary a Lorentzian signature fixed radial slice $\Sigma_r$ with boundary conditions preserving the conformal group. Naively, one might expect to find Lorentzian Liouville theory. This turns out not to be the case. The Hamiltonian reduction is \emph{in fine} independent of the chosen radial slice. Since a fixed radial slice close to $\mathcal I^+$ leads to the Euclidean Liouville theory, the same theory is found on a fixed radial slice inside the static patch, namely
\bea
S_{EH} = -\frac{\ell^2}{64 \pi G} \int d\phi \, dt \left(  \left(\p_t \Phi\right)^2 + \frac{1}{\ell^2} \left(\p_\phi \Phi\right)^2 +\frac{16}{\ell^2} e^\Phi \right)\, ,
\eea
where the boundary terms of the Einstein-Hilbert action $S_{EH}$ were chosen to enforce the boundary conditions. The awkward feature is now that $t$, the Euclidean time in the boundary field theory, is a timelike coordinate of  the boundary $\Sigma_r$. Overall, our result is consistent with the dS/CFT conjecture \cite{Strominger:2001pn}: we find  a Euclidean CFT, even when the holographic boundary is a timelike cylinder in the static patch. Note that there is no holographic RG flow in the sense of \cite{Heemskerk:2010hk} since no bulk fields are integrated out upon displacing the holographic boundary into the bulk. 

Our derivation can be extended in a straightforward manner to higher spin fields as long as no propagating degrees of freedom are involved. We expect that the notion of asymptotic symmetry can be realized everywhere in the bulk and we similarly expect that the Hamiltonian reduction can be done on any slice in the bulk without any dependence on the choice of slice. The addition of propagating modes on the other hand is non-trivial and further analysis would be required. 

On the technical side, we use the reformulation of Einstein gravity with positive cosmological constant as two copies of $SL(2,\mathbb C)$ Chern-Simons theory with a reality constraint  \cite{Achucarro:1987vz,Witten:1988hc}. We note that the Fefferman-Graham gauge for the metric naturally leads to the highest weight gauge for the first Chern-Simons gauge field and lowest weight gauge for the second. Instead, Eddington-Finkelstein coordinates for the metric, which cover both the conformal boundary and the static observer, lead to a highest weight gauge for both Chern-Simons gauge fields. This distinction leads to some new features of the Hamiltonian reduction to Liouville theory with respect to previous treatments \cite{Coussaert:1995zp,Banados:1998pi,Cacciatori:2001un,Forgacs:1989ac}.  Usually, one performs a Gauss decomposition of an $SL(2,\mathbb C)$ element around the identity in order to reduce the non-chiral WZW model to Liouville theory. Here, it turns out that a natural Gauss decomposition involves particular coordinates far from the identity, in order to parameterize the Liouville field without otherwise intricate field redefinitions. 

The rest of the paper is organized as follows. In Section 2, we derive the  symmetry algebra of pure Einstein gravity in the bulk spacetime, both at the level of asymptotic Killing vector fields and associated conserved charges. In Section 3, we review the Chern-Simons formalism for asymptotically $dS_3$ spacetimes and present the classical phase space of spinning conical defects equipped with Virasoro gravitons in two sets of coordinates of interest. We perform the reduction to the WZW model and then to Liouville theory in Section 4. Our conventions are given in appendix.

\section{Asymptotic symmetries everywhere}

The phase space of Einstein gravity with positive cosmological constant in three dimensions can be written  in Eddington-Finkelstein gauge as
\begin{eqnarray}
ds^2=\left( \frac{r^2}{l^2}+8G \rd M(u,\phi)\right)du^2-2dudr+8G \rd J(u,\phi) du d\phi+r^2d\phi^2,
\label{metricBMS}
\end{eqnarray}
where the functions $\rd M(u,\phi)$, $\rd J(u,\phi)$ satisfy
$
\partial_u \rd J= \partial_\phi \rd M$ and $\partial_u \rd M=-\frac{1}{l^2}\partial_\phi \rd J$. Note that we will keep all factors of $\ell$ explicit in order to also discuss the AdS analytic continuation $\ell \rightarrow i \ell$ and the flat spacetime limit $\ell \rightarrow \infty$. 

\subsection{Symmetry algebra}

The phase space is preserved under the action of the vector field
\begin{align}\label{bms3:438}
\xi=f\partial_u&+\left(-r\partial_\phi Y+\partial_\phi^2 f-\dfrac{8G\rd J}{2r}\partial_\phi f\right)\partial_r+\left(Y-\dfrac{\partial_\phi f}{r}\right)\partial_\phi,
\end{align}
where the functions $f(u,\phi)$ and $Y(u,\phi)$ satisfy 
$\partial_u f=\partial_\phi Y$, $\partial_u Y=-\frac{1}{l^2}\partial_\phi f$.
Interestingly, the perturbative expansion in $r$  of the symmetry generator in this gauge stops at next-to-next-to-leading order.

At leading order close to future infinity $\mathcal I^+$ (defined as the limit $r\to \infty$), the vector field \eqref{bms3:438} reduces to
\begin{eqnarray}\label{5:bms}
\bar\xi=f\partial_u- r \partial_\phi Y \partial_r +Y\partial_\phi,
\end{eqnarray}
and its algebra is found to be
\begin{eqnarray}\label{bms3:algebra}
[\bar\xi_1,\bar\xi_2]\equiv\hat f\partial_u - r \partial_\phi \hat Y \partial_r +\hat Y\partial_\phi,
\end{eqnarray}
where
\begin{eqnarray}
\badat{2}
\hat f&=Y_1\partial_\phi f_2-Y_2\partial_\phi f_1+f_1\partial_\phi Y_2-f_2\partial_\phi Y_1,\\
\hat Y&=Y_1\partial_\phi Y_2-Y_2\partial_\phi Y_1-\dfrac{1}{\ell^2}\left(f_1\partial_\phi f_2-f_2\partial_\phi f_1\right).
\label{bms3:421}
\eadat
\end{eqnarray}
These relations define the symmetry algebra and can be written more compactly as
\begin{align}\label{bms3:al}
[(f_1,Y_1),(f_2,Y_2)]=(\hat f,\hat Y).
\end{align}

When $\ell$ is finite, it is convenient to define the coordinates $t^\pm=u \pm i \ell \phi$. One has $\partial_+\partial_-f=0=\partial_+\partial_-Y$, which can be integrated for $f,Y$ in terms of two arbitrary functions $l^+(t^+)$, $l^-(t^-)$:
\begin{eqnarray}
f=\frac{1}{2}\left(l^++l^-\right),\hspace{1cm}Y=\dfrac{-i}{2\ell}\left(l^+-l^-\right).
\end{eqnarray}
The leading-order symmetry vector \eqref{5:bms} therefore becomes $ \bar\xi =l^+\partial_+~+~l^-\partial_- - \frac{ r}{2} (\p_+ l^+ + \p_- l^- ) \partial_r$ and, expanding the generators as
\begin{eqnarray}
\badat{2}
l_m^+=\{\bar\xi:\hspace{0.5cm}l^+=\ell e^{-m\frac{t^+}{\ell}},l^-=0\}= \ell e^{-m\frac{t^+}{\ell}}\left( \partial_+ + \frac{m}{2l}r \p_r\right) ,\\
l_m^-=\{\bar\xi:\hspace{0.5cm}l^+=0,l^-=\ell e^{-m\frac{t^-}{\ell}}\}= \ell e^{-m\frac{t^-}{\ell}} \left( \partial_-  + \frac{m}{2l}r \p_r\right),
\eadat
\end{eqnarray}
one finds that the algebra of the vector fields consists of two copies of the Witt algebra
\begin{eqnarray}
[l_m^\pm,l_m^\pm]=(m-n)l_{m+n}^\pm.
\end{eqnarray}
Note the relations $(l^\pm_m)^*=l^\mp_{m}$.

\subsubsection*{Modified Lie bracket and symmetry realization in the bulk}

The bulk symmetry parameter \eqref{bms3:438} is field dependent (through the metric function $\mathcal J$) and therefore its algebra is given by the modified bracket \cite{Barnich:2010eb,Barnich:2010xq}
\begin{eqnarray}\label{bms3:m}
[\xi_1,\xi_2]_M=[\xi_1,\xi_2]+\delta_{\xi_1}\xi_2(g)-\delta_{\xi_2}\xi_1(g).
\end{eqnarray}
The rationale for this definition is as follows. The $\p_\mu$ derivative in the commutator acts on the fields appearing in the symmetry parameters. These contributions are then canceled by the two additional terms. The signs are fixed with the convention $\delta_\xi g_{\mu\nu} = -\mathcal L_\xi g_{\mu\nu}$. 

By means of this modified bracket, one can show that the bulk field \eqref{bms3:438} forms a representation of the symmetry algebra \eqref{bms3:al}:
\begin{align}
[\xi_1,\xi_2]_M=
\hat f\partial_u&+\left(-r\partial_\phi \hat Y+\partial_\phi^2 \hat f-\dfrac{8G\rd J}{2r}\partial_\phi \hat f\right)\partial_r+\left(\hat Y-\dfrac{\partial_\phi \hat f}{r}\right)\partial_\phi.
\end{align}
The symmetry algebra is thus represented everywhere in the bulk of the spacetime even though it has been defined at infinity.

\subsection{Surface charge algebra}

The surface charges associated with the symmetry generator \eqref{bms3:438} are now computed using the covariant formalism \cite{Barnich:2001jy,Barnich:2007bf}.
A 1-form $\ndelta \mathcal Q_\xi$
\footnote{The charge can be non integrable, hence the $\ndelta$ notation.}
which depends on a solution $g$ and its variation $\delta g$ is associated to a vector field $\xi$. $\ndelta \mathcal Q_\xi$ is defined in $n$ spacetime dimension
by
\begin{eqnarray}
\ndelta Q_\xi [\delta g, g]=\frac{1}{8 \pi G} \int_{\partial \Sigma}(d^{n-2} x)_{\mu \nu}\,  \sqrt{-g}~~ \,
\left(\xi^\nu D^\mu \delta g-\xi^\nu D_\sigma \delta g^{\mu \sigma}+\xi_\sigma D^\nu \delta g^{\mu \sigma}+\frac{1}{2}\delta g D^\nu \xi^\mu+\notag\right.\\
\left.+\frac{1}{2}\delta g^{\nu \sigma}(D^\mu \xi_\sigma-D_\sigma \xi^\mu) \right) ,
\label{formulacharge}
\end{eqnarray}
where 
$(d^{n-2} x)_{\mu \nu}\equiv \frac{1}{2! (n-2)!} \ep_{\mu \nu \sigma_1 \cdots \sigma_{n-2}} dx^{\sigma_1}\wedge \cdots \wedge dx^{\sigma_{n-2}}$ denotes the dual of a $2$-form in $n$ dimensions.\\

In three dimensions and with $g$ given by \eqref{metricBMS}, the surface integration $\partial \Sigma$ is taken to be the circle ($u$ and $r$ fixed) and one finds that
\begin{align}
\ndelta\mathcal Q_{\xi}(g,\delta g)&=\dfrac{1}{2\pi }\int_0^{2 \pi} \left(
f\delta \rd M+Y\delta \rd J-\dfrac{1}{2r}{(f\partial_\phi\delta \rd J+\delta \rd J\partial_\phi f)}
\right) d\phi .
\label{bms3:4:28}
\end{align}

Crucially, the $1/r$ term vanishes due to an integration by parts with respect to the $\phi$ coordinate. Because the remaining right-hand side of \eqref{bms3:4:28} is made of $\delta$-exact terms, the associated charge is integrable and reads
\begin{eqnarray}
\mathcal Q_{\xi}~=~\dfrac{1}{2\pi}\int_0^{2 \pi} \left(f \rd M+Y \rd J\right)d\phi.
\label{bms3:4:29}
\end{eqnarray}
Here, we fixed the normalization such that $\mathcal Q_{\xi}$ is  zero for  $\mathcal M= \mathcal J =0$. 
The charge is $r$ independent. Therefore, this expression for the charge is the same everywhere in the bulk of the spacetime.

One could have also used the Iyer-Wald formula for the charges \cite{IyerWald}, which is equal to the expression \eqref{formulacharge} with the last term removed. The final term might in general be non-zero for non-Killing vectors fields, such as the symmetries that we are using. However, the term evaluates to zero, and the Iyer-Wald charges are identical to \eqref{bms3:4:29}. 

The charge formula \eqref{bms3:4:29} makes explicit the relationship between the integration functions of the  symmetries ($f,Y$) and the integration functions of the solution to the equations of motion ($\rd M, \rd J$). More precisely, the charge $\mathcal Q_\xi$ in \eqref{bms3:4:29} provides an inner product between the space of solutions and the asymptotic symmetries.

Upon defining $\mathcal M  = \mathcal L_+ + \mathcal L_-$, $\mathcal J = i\ell( \mathcal L_+ - \mathcal L_-)$, the charge is given by
\begin{eqnarray}
\mathcal Q_{\xi}~=~\dfrac{1}{2\pi}\int_0^{2 \pi} \left( l^+ \mathcal L_+ + l^- \mathcal L_- \right)d\phi,
\end{eqnarray}
which also makes manisfest the relationship between the functions $(l^+,l^-)$ and the integrations functions of the solution $(\mathcal L_+,\mathcal L_-)$. Note that the semi-classical spectrum of $\rd L_+$ and $\rd L_-$ is complex.

It is worth pointing out that the result \eqref{bms3:4:29} is valid for asymptotically flat, anti-de Sitter and de Sitter cases. The anti-de Sitter case is simply obtained by analytic continuation $\ell \rightarrow i \ell$. The asymptotically flat case is then obtained by taking the limit $\ell \rightarrow \infty$. Since all quantities $f,Y,\mathcal M,\mathcal J$ are finite in the flat limit, one readily obtains the result. (One cannot however use $l^\pm_m$ which are not well-defined in the flat limit). 

More conceptually, the fact that the charges are independent of the radius follows from the vanishing of the symplectic structure of the theory. Indeed, the symplectic structure evaluated on the Lie derivative of the metric is a boundary term, $\omega(\mathcal L_\xi g_{\mu\nu},\delta g_{\mu\nu},g) = d k_\xi (\delta g,g)$ where $\ndelta Q_\xi [\delta g, g] = \int k_\xi(\delta g,g)$  is precisely the charge \eqref{formulacharge}. The vanishing of the symplectic structure implies that  the difference of charge $\ndelta Q_\xi$ evaluated on two surfaces  $r=r_1$ and $r=r_2$ constant is zero. Therefore, the charge is independent of the radius.

\subsubsection*{Algebra of surface charges: two Virasoro in the bulk}

The transformation laws of the functions  $\mathcal M, \mathcal J$ under the symmetry transformation generated by \eqref{bms3:438} are given by 
\begin{align}
\badat{2}
-\delta \rd M&=Y\partial_\phi \rd M+2\rd M\partial_\phi Y -\frac{1}{4G}\partial_\phi^3Y-\dfrac{1}{l^2}(2\rd J\partial_\phi f+f\partial_\phi \rd J) ,\\
-\delta \rd J&=Y\partial_\phi \rd J+2 \rd J \partial_\phi Y -\frac{1}{4G}\partial_\phi^3f+2\rd M\partial_\phi f +f \partial_\phi \rd M.
\eadat
\end{align}

One can rewrite the transformation laws as 
\begin{align}
-\delta \mathcal L_\pm = l_\pm \p_\pm \mathcal L_\pm + 2 \mathcal L_\pm \p_\pm l_\pm + \frac{\ell^2}{8G} \p_\pm^3 l_\pm .
\end{align}

The algebra of surface charges \eqref{bms3:4:29} can then be computed with the Poisson bracket defined by
\begin{eqnarray}
\{\mathcal Q_{\xi_1},\mathcal Q_{\xi_2}\}=\delta_{\xi_1}\mathcal Q_{\xi_2}.
\end{eqnarray}
One finds
\begin{align}
\delta_{\xi_1}\mathcal Q_{\xi_2}
&=\dfrac{1}{2\pi }\int_0^{2 \pi} d\phi\left(\hat f \rd M+\hat Y \rd J\right)-\dfrac{1}{8\pi G}\int_0^{2 \pi}d\phi (f_1\partial_\phi^3Y_2+Y_1 \partial_\phi^3f_2).\label{bms3:462}
\end{align}
Therefore, one has
\begin{eqnarray}\label{bms3:463}
\{\mathcal Q_{\xi_1},\mathcal Q_{\xi_2}\}=\mathcal Q_{[\xi_1,\xi_2]}+\mathcal K_{\xi_1,\xi_2},
\end{eqnarray}
where $\mathcal K_{\xi_1,\xi_2}$ is by definition the second term of \eqref{bms3:462}.

Introducing $L_m^\pm=\mathcal Q_{l^\pm_m}$, we find that the charge algebra consists of two copies of the Virasoro algebra 
\begin{align}\label{vira}
\{L_m^\pm,L_n^\pm\}=(m-n)L_{m+n}^\pm+\dfrac{c^\pm}{12}m^3\delta_{m+n,0},
\end{align}
everywhere in the bulk, with central charge $c^\pm=\frac{3l}{2G}$. The charges obey $(L_m^+)^*=L^-_m$. Note that with our definitions there is no $i$ on the left-hand side of the above relation \eqref{vira}. This is in contrast to  the AdS result  \cite{Brown:1986nw}. 

In the AdS case obtained by analytical continuation, it similarly follows that the Brown-Henneaux realization of asymptotic symmetries \cite{Brown:1986nw} can be extended everywhere in the bulk. In the case of the asymptotically flat limit, we have shown that the $bms_3$ charge algebra \cite{Barnich:2006av}  is defined everywhere into the bulk. All these results are valid for three-dimensional Einstein gravity without matter. The generalization with propagating modes is far from obvious.

\section{Chern-Simons formulation}

Three-dimensional Einstein gravity with positive cosmological constant can be formulated as two copies of $SL(2,\mathbb C)$ Chern-Simons theory with a reality constraint \cite{Achucarro:1987vz, Witten:1988hc}, with the action 
\begin{equation}\label{EHaction}
S_E[A,\bar A] =-i S_k[A] + i S_k[\bar A] = \frac{1}{16\pi G_3}\int_{Bulk}  d^3x\, \sqrt{-g}\left(R-\frac{2}{\ell^2}\right) +\text{boundary term}\, ,
\end{equation}
where $k = \ell/(4G)$ and
\bea
S_k[A] &=& \frac{k}{4\pi}\int_{Bulk}\, {\rm Tr}\left(A\wedge dA+\frac{2}{3}A\wedge A\wedge A\right) +
\text{boundary term}\, .
\eea
The equations of motion are given by 
\bea
F \equiv dA + A \wedge A =0 \, ,\qquad \ \bar F \equiv d\bar A + \bar A \wedge \bar A =0 ,
\eea
where 
\begin{eqnarray}
A=A^{a}\tau_{a}=\left(\omega^{a}+\frac{i}{\ell}e^{a}\right)\tau_{a},\qquad \bar A=\bar A^{a}\tau_{a}=\left(\omega^{a}-\frac{i}{\ell}e^{a}\right)\tau_{a}  
\end{eqnarray}
and $\omega^{a} = \frac{1}{2}\eps^{a b c}\omega_{b c}$. Here, $\tau_{a}$ are $SL(2,\mathbb C)$ generators which are normalized as
$ 
{\rm Tr}(\tau_{a} \tau_{b}) = \frac{1}{2}\eta_{a b} 
$.

We consider Dirichlet boundary conditions, which we will present in Section \ref{bndterms}. The resulting classical phase space contains spinning conical defects studied in 1984 by Deser and Jackiw \cite{Deser:1983nh}. It also contains the ``boundary gravitons'' or Virasoro descendants which were derived in \cite{Strominger:2001pn}. We will here present the space of solutions in two distinct coordinate systems which have distinct features. Fefferman-Graham coordinates are adapted to the conformal boundary and its holographic interpretation in terms of a CFT. However, already for the vacuum, these coordinates do not cover the static patch since they break down at the cosmological horizon. In contrast, Eddington-Finkelstein coordinates cover both the future diamond and static patch of global de Sitter. 

\subsection{Fefferman-Graham slicing}\label{FGslicing}

We consider asymptotically de Sitter metrics of the form
\begin{align}
ds^2 = -\ell^2 \frac{d\tau^2}{\tau^2} + \left(\tau^2 + \frac{16 G^2 {\cal L}_+\left(t^+\right) {\cal L}_-(t^-)}{\tau^2}\right)dt^+ dt^- - 4 G \, {\cal L}_+(t^+) \ ( dt^+)^2  - 4 G \, {\cal L}_-(t^-)\  (dt^-)^2\, ,
\end{align}
where $t^\pm = t \pm i \ell \phi$, $\phi \sim \phi + 2\pi$\footnote{For book-keeping purposes, we have explicitly $t=\frac{1}{2}(t^+ + t^-)$, $\phi = \frac{i}{2\ell}(-t^+ + t^-)$, $\p_t = \p_+ + \p_ -$, $\p_\phi = i \ell (\p_+ - \p_-)$, $\p_+ = \frac{1}{2} (\p_t - \frac{i}{\ell} \p_\phi)$, $\p_- = \frac{1}{2} (\p_t + \frac{i}{\ell} \p_\phi)$.}.
The complex functions ${\cal L}_\pm$ parametrize the phase space of such metrics. They are constrained by the relation ${\cal L}_+^*={\cal L}_-$. It is convenient to define the real functions
\begin{equation}
\mathcal M(t^+,t^-)={\cal L}_+(t^+)+{\cal L}_-(t^-) \ ,\qquad \mathcal J(t^+,t^-)=i \ell ( {\cal L}_+(t^+)-{\cal L}_-(t^-)) \,
\end{equation}
which zero modes are the mass and angular momentum. The coordinate system breaks down at $\tau=0$ or even at the larger $\tau= 2 G^{1/2} ({\cal L}_+ {\cal L}_-)^{1/4}$ if ${\cal L}_+{\cal L}_- >0$. 

\begin{figure}
\begin{minipage}{0.465\textwidth}
\centering
\def\svgwidth{1.1\textwidth}
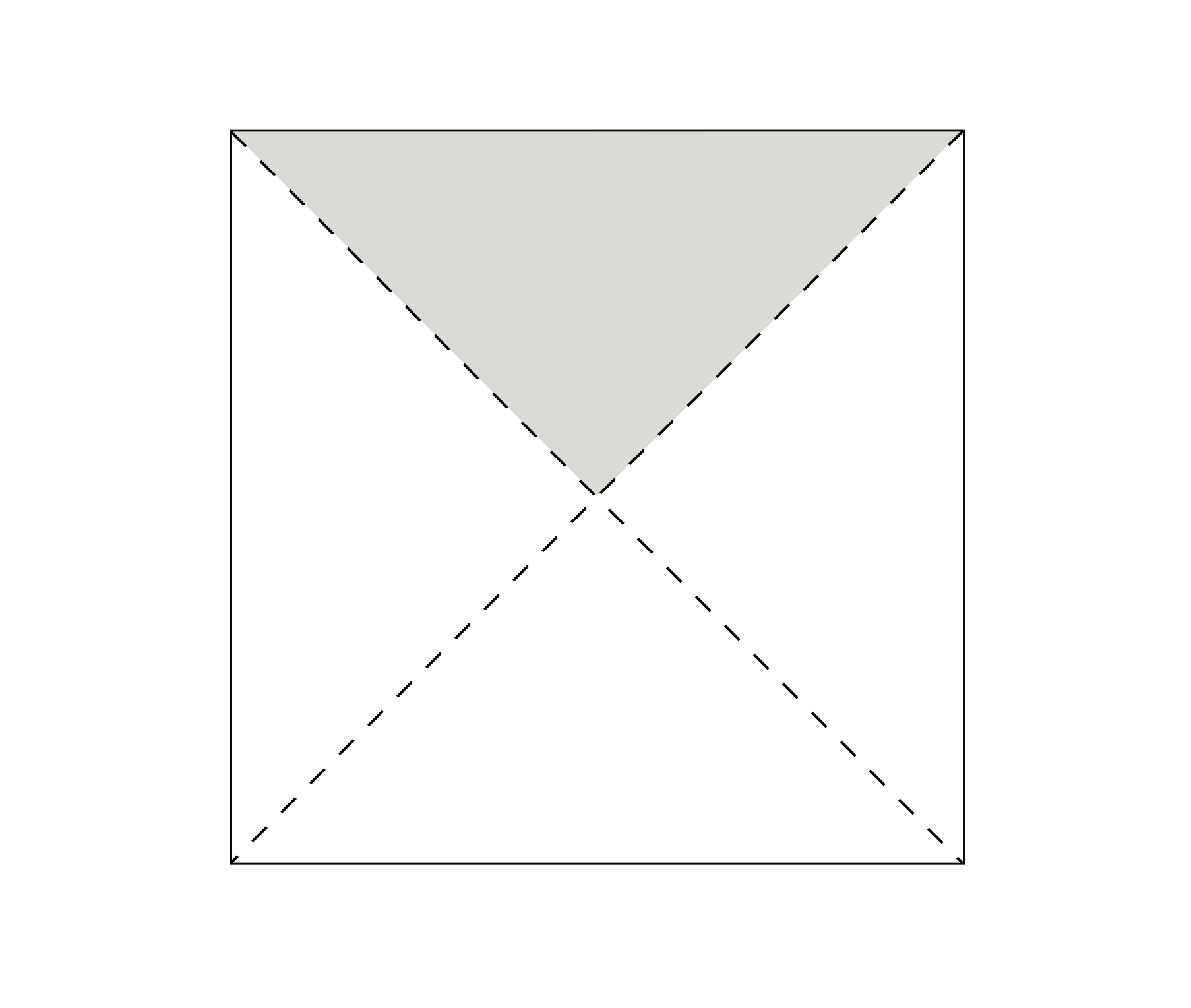
\caption{Fefferman-Graham coordinates}
\label{FGfigure}
\end{minipage}
\begin{minipage}{0.465\textwidth}
\centering
\def\svgwidth{1.1\textwidth}
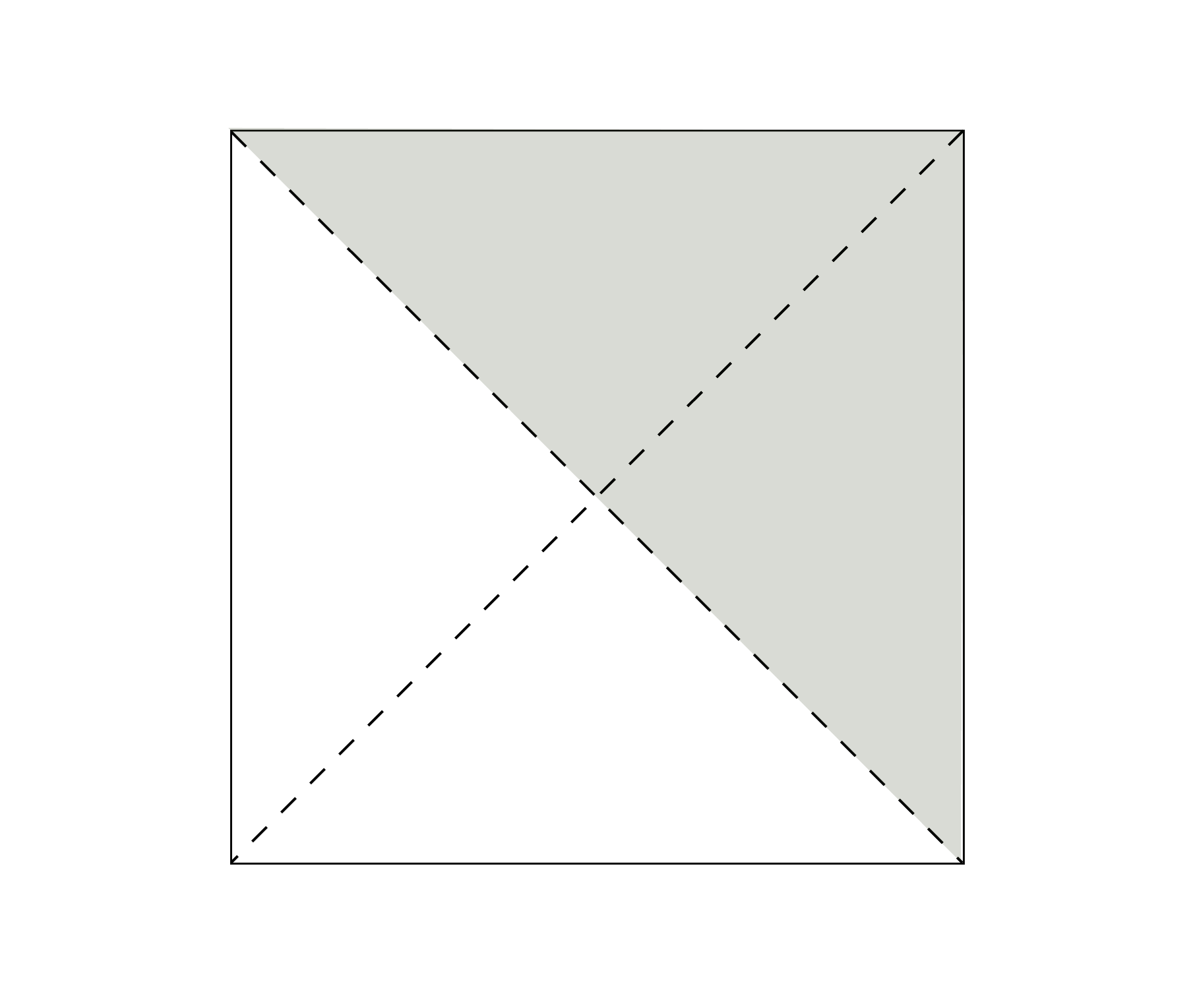
\caption{Eddington-Finkelstein coordinates}
\label{EFfigure}
\end{minipage}
\end{figure}

This coordinate system is not suitable to describe  the coordinate patch of the static observer at the south pole beyond his cosmological horizon. To see this, let us consider the  case of the $dS_3$ vacuum, with $\mathcal M = \frac{1}{8G}$, $\mathcal J=0$: 
\bea
ds^2  = -\ell^2 \frac{d\tau^2}{\tau^2} + \left(\tau- \frac{1}{4\tau}\right)^2 dt^2 + \ell^2 \left(\tau+ \frac{1}{4\tau}\right)^2 d\phi^2
\eea
which is valid when $\frac{1}{2} \leq \tau \leq \infty$. One recognizes the static patch coordinates after defining $r = \tau+\frac{1}{4\tau}$, $1 \leq r \leq \infty$ such that 
\bea
ds^2  = -\ell^2 \frac{dr^2}{r^2-1} + (r^2-1) dt^2 + \ell^2  r^2 d\phi^2\, .
\eea
This coordinate system only covers the upper diamond of global de Sitter, see figures \ref{FGfigure} and \ref{EFfigure}.

To obtain the gauge field we have to specify the vielbein and the $SL(2,\mathbb C)$ generators. The choice of the $SL(2,\mathbb C)$ generators, $\tau_a$, should be consistent with $ds^2=\eta_{ab}e^ae^b$ and ${{\rm Tr} \, \tau_{a}\tau_{b}}=\frac{1}{2}\eta_{ab}$. We choose
\bea
e^0 &=&- \frac{\ell}{r}dr, \nonumber\\
e^1 &=& -r \, dt +\frac{2G}{r} \left(\mathcal M \, dt + \mathcal J \, d\phi\right),\nonumber\\
e^2 &=& -\ell r\,  d\phi - \frac{2 \ell G  }{r} \left(\mathcal M \, d\phi - \frac{\mathcal J}{\ell}dt\right) ,
\eea
\bea
\tau^{FG}_0 = -i L_0,\qquad
\tau^{FG}_1 = \frac{1}{2}(L_1 - L_{-1}),\qquad
\tau^{FG}_2 = \frac{i}{2}(L_1 + L_{-1})
\eea
where $L_{\pm 1},L_0$ are defined in the appendix.  We then have $\eta_{ab}=\text{diag}(-1,1,1)$. The gauge fields $A$ and $\bar A$ are then
\begin{align}
A^{FG} &=\frac{1}{2r}  \left(  \begin{array}{cc} 
1 & 0 \\
0& -1
 \end{array} \right)dr+ \left(  \begin{array}{cc} 
0 & \frac{4 i G {\cal L}_+(t^+)}{\ell r}\\
-\frac{i r}{\ell}  &0
 \end{array} \right)dt^+  \, ,
 \\
 \bar A^{FG}& = 
\frac{1}{2r} \left(  \begin{array}{cc} 
-1 & 0 \\
0 & 1
 \end{array} \right) dr
 +
 \left(  \begin{array}{cc} 
0 & \frac{ir}{\ell} \\
-\frac{4 i G {\cal L}_-(t^-)}{\ell r}  & 0
 \end{array} \right) dt^-\ . 
 \end{align}
In the approach of \cite{Coussaert:1995zp}, the boundary conditions are specified at future infinity for the gauge field after the $r$-dependence is factorized out. An interesting feature of de Sitter space time  in the  Fefferman-Graham coordinates is that the $r$-dependence factorizes out not only at the future infinity but in the whole upper diamond of the Penrose diagram. We call the  $r$-independent factor of the gauge field the reduced gauge connection $a^{FG}$:
\bea
a^{FG} &= &  -\frac{i}{\ell} \left(  \begin{array}{cc} 
0& -4 G {\mathcal L_+(t^+)}  \\
1  &0
 \end{array} \right)dt^+ = -i \left(\frac{1}{\ell} L_1 + \frac{1}{k}{\mathcal L_+(t^+)} { L}_{-1}\right) dt^+, \nonumber\\ 
\bar  a^{FG} &=&  \frac{i}{\ell} \left(  \begin{array}{cc} 
0& 1 \\
- 4 G {\mathcal L_-(t^-)}  &0
 \end{array} \right)dt^- = -i \left( \frac{1}{\ell} {L}_{-1} +\frac{1}{k} {\mathcal L_-(t^-)} L_1 \right)dt^-\, ,
\eea
where $A^{FG},\bar A^{FG}$ and $a^{FG},\bar a^{FG}$ are related by the gauge transformation
\begin{equation}\label{KFG}
a^{FG} = K^{-1} A^{FG} K + K^{-1} d K\, ,\qquad  \bar a^{FG} = K \bar A^{FG} K^{-1} + K d K^{-1}\ , 
\end{equation}
with $K = \text{diag}(r^{-1/2},r^{1/2})$. 

A  useful property of this basis is
\bea\label{defsigma}
(\tau_a^{FG})^\dagger = \sigma  \tau^{FG}_a \sigma,\qquad {\rm with}\qquad \sigma \equiv 2i L_0 =\left( \begin{array}{cc} i & 0 \\0 &-i\end{array}\right) \, .\label{rel}
\eea
The reduced gauge connection $a^{FG}$ is in lowest weight form while $\bar a^{FG}$ is in highest weight form.  As a consequence of \eqref{rel}  they are related by
\begin{equation}
a_{FG}^\dagger = \sigma \bar a_{FG} \sigma = \bar a_{FG} .
\end{equation}

\subsection{Eddington-Finkelstein slicing}\label{EFslicing}

Since Fefferman-Graham coordinates do break at the cosmological horizon, it is necessary to consider another coordinate system in order to impose boundary conditions beyond the horizon. We will now repeat all steps from the previous subsection in Eddington-Finkelstein type coordinates.

The phase space of asymptotically de Sitter spacetimes is now given by
\bea
ds^2 = \left(\frac{r^2}{\ell^2}-8  G ({\cal L}_++{\cal L}_-) \right)du^2 -2du dr - 8 i\ell\, G ({\cal L}_+-{\cal L}_-) du d\phi +r^2 d\phi^2 
\eea
where $u \in \mathbb R$, $\phi \sim \phi+2\pi$ and $0 \leq r$. 

We choose
\begin{align}
e^0 = \left(\frac{r^2}{2\ell^2}-4G\,({\cal L}_++{\cal L}_-) \right)du-dr-4i\ell G\,  ({\cal L}_+-{\cal L}_-)d\phi,\qquad &e^1=-2du,\qquad e^2=rd\phi,\nonumber\\
\tau^{EF}_0 =-\frac{1}{2}L_1,\qquad \tau^{EF}_1 = -\frac{1}{2}L_{-1},\qquad \tau^{EF}_2 = L_0,
\end{align}
such that $ds^2 = -e^0 e^1 + (e^2)^2 \equiv \eta_{ab}e^a e^b$. 

From the choice of the generators and dreibein,  we obtain the gauge fields
\begin{align}
A^{EF}&= \frac{i}{2\ell}\left(\begin{matrix}0 &0\\ 1&0\end{matrix}\right)dr
+\left(
\begin{matrix}
\frac{r}{2\ell^2} & -\frac{i}{\ell}\\ -\frac{ir^2}{4\ell^3}+\frac{i \mathcal L_+}{ k}&-\frac{r}{2\ell^2}
\end{matrix}
\right)dt^+\, ,\nonumber\\
\bar A^{EF}&=- \frac{i}{2\ell}\left(\begin{matrix}0 &0\\ 1&0\end{matrix}\right)dr
+\left(
\begin{matrix}
\frac{r}{2\ell^2} & \frac{i}{\ell}\\ \frac{ir^2}{4\ell^3}-\frac{i \mathcal L_-}{k }&-\frac{r}{2\ell^2}
\end{matrix}
\right)dt^-\, ,
\end{align}
where we defined $t^\pm = u \pm i \ell \phi$.

As in the case of the Fefferman-Graham coordinates the $A_-$ and $\bar A_+$ contributions are zero. 
Again, it is possible to factorize out the r-dependence.
We define the reduced gauge connection as 
\bea
a^{EF} = K^{-1} A^{EF} K +K^{-1}dK, \qquad \bar a^{EF} = K \bar A^{EF} K^{-1} +KdK^{-1}
\eea
where
$
K = \left(  \begin{array}{cc} 1 & 0 \\ -\frac{i}{2\ell} r  & 1 \end{array} \right).
$
Note that the form of the matrix $K$ differs from the one  in Fefferman-Graham coordinates \eqref{KFG}.

On-shell we find for the reduced gauge field
\bea
a^{EF} &=&   \left(  \begin{array}{cc} 
0& -\frac{i}{\ell}  \\ 
\frac{i {\cal L}_+(t^+)}{k } &0 
 \end{array} \right) dt^+ =\frac{i}{\ell} \left(L_{-1}+\frac{\ell}{k}{\cal L}_+(t^+) L_1\right) dt^+ \label{defa} , \\
  \bar a^{EF} &=&   \left(  \begin{array}{cc} 
0& \frac{i}{\ell}  \\ 
\frac{-i {\cal L}_-(t^-)}{k } &0 
 \end{array} \right) dt^-= -\frac{i}{\ell} \left( L_{-1}+\frac{\ell}{k}{\cal L}_-(t^-) L_1\right) dt^-.
\eea

Since the basis of generators $\tau^{EF}_a$ is real, it implies $\bar a_{EF}= a_{EF }^*$. 
A  useful property of this basis is
\bea\label{defsigmahat}
(\tau_a^{EF})^\dagger =- \hat{\sigma} \tau^{EF}_a \hat{\sigma},\qquad {\rm with}\qquad\hat\sigma \equiv i( L_1 + L_{-1})=\left( \begin{array}{cc} 0&-i\\i&0\end{array}\right) \, .
\eea

\subsection{Boundary conditions}\label{bndterms}

Let us define a slicing of (a part of) spacetime into fixed radial slices $\Sigma_r$, such that in the limit $r \rightarrow \infty$, $\Sigma_\infty$ coincides with the future conformal boundary ${\mathcal I}^+$. There is an infinite number of such slicings. Two examples (Fefferman-Graham and Eddington-Finkelstein slicings) were provided above.  We then define the reduced gauge connections $a$ and $\bar a$ as
\bea
a = K^{-1} A K + K^{-1} d K,\qquad \bar a = \bar K^{-1} \bar A  \bar K  +\bar K^{-1} d \bar K \label{defK}
\eea
such that $a_r = 0 = \bar a_r$. This fixes $K, \, \bar K \in SL(2,\mathbb C)$ up to an $SL(2,\mathbb C)$ element  on $\Sigma_r$ which corresponds to $(t^+,t^-)$-dependent diffeomorphisms tangent to the slices. For simplicity, we will assume that $K$ and $\bar K$ only depend on $r$. 

We are now ready to state our boundary conditions. They come in two sets:
\begin{enumerate}
\item $A_- = \bar A_+ = 0$ on $\Sigma_r$.
\item $a_+ =  \frac{i}{\ell} L_{-1} + 0\, L_0 +{\cal O}(1) L_1$ and $\bar a_- =  - \frac{i}{\ell} L_{-1} +0 \, L_0 +{\cal O}(1) L_1$ on $\Sigma_r$, where $L_{-1},L_0,L_1$ form the canonical $SL(2,\mathbb R)$ algebra given in the appendix. 
\end{enumerate}

The phase space in Eddington-Finkelstein coordinates clearly obeys the boundary conditions, with $K=\bar K^{-1}$ given above. In fact, the phase space in Fefferman-Graham coordinates also obeys the boundary conditions, once we realize that the definition of $SL(2,\mathbb R)$ generators in the boundary conditions is related to the choice of generators in \eqref{defa} via the inner automorphism $\hat \sigma $ of the algebra defined in appendix. More precisely, we have the following relationship between the reduced connections obtained from Fefferman-Graham  and Eddington-Finkelstein coordinates (and our choice of basis and dreibein):
\bea
a^{EF} =   \hat \sigma^{-1} a^{FG} \hat \sigma   ,\qquad \bar a^{EF} = \bar a^{FG}\, .\label{FGa}
\eea
Therefore, for $\Sigma_\infty$, $K=\bar K^{-1} = \text{diag}(r^{-1/2},r^{1/2})$ and after applying the automorphism on the $a$ sector, the boundary conditions exactly coincide with the ones of \cite{Cacciatori:2001un}.

Note that in the two phase spaces that we considered, one has $\p_- A_+ = 0$ and $\p_+ \bar A_+ = 0$. These conditions are not part of the boundary conditions but are only on-shell conditions. 

\section{Hamiltonian reduction}

\begin{figure}
\begin{minipage}{0.465\textwidth}
\centering
\def\svgwidth{1.1\textwidth}
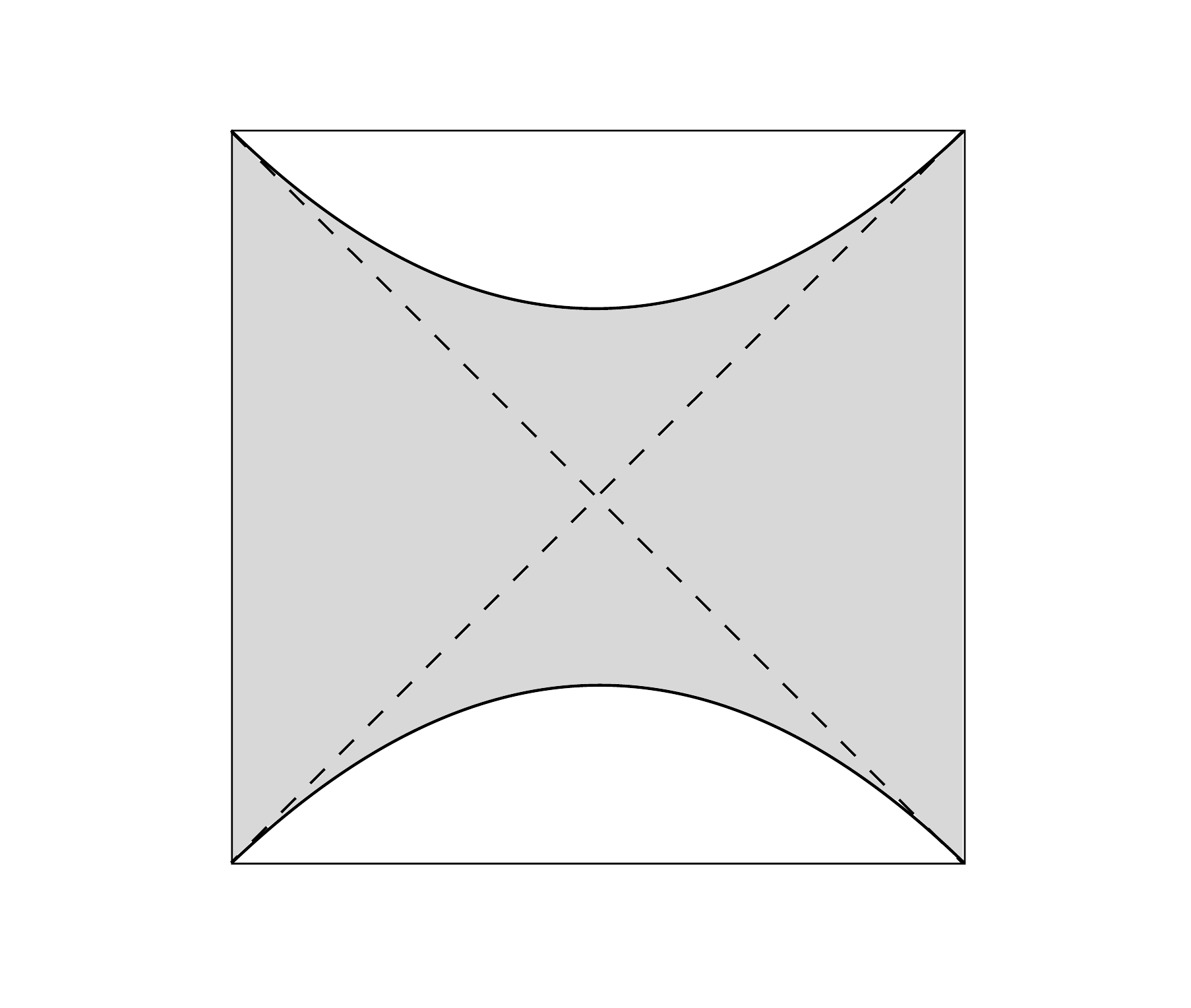
\caption{Bulk region bounded by $\Sigma_r^+$ and $\Sigma_r^-$}
\label{spacelike}
\end{minipage}
\begin{minipage}{0.465\textwidth}
\centering
\def\svgwidth{1.1\textwidth}
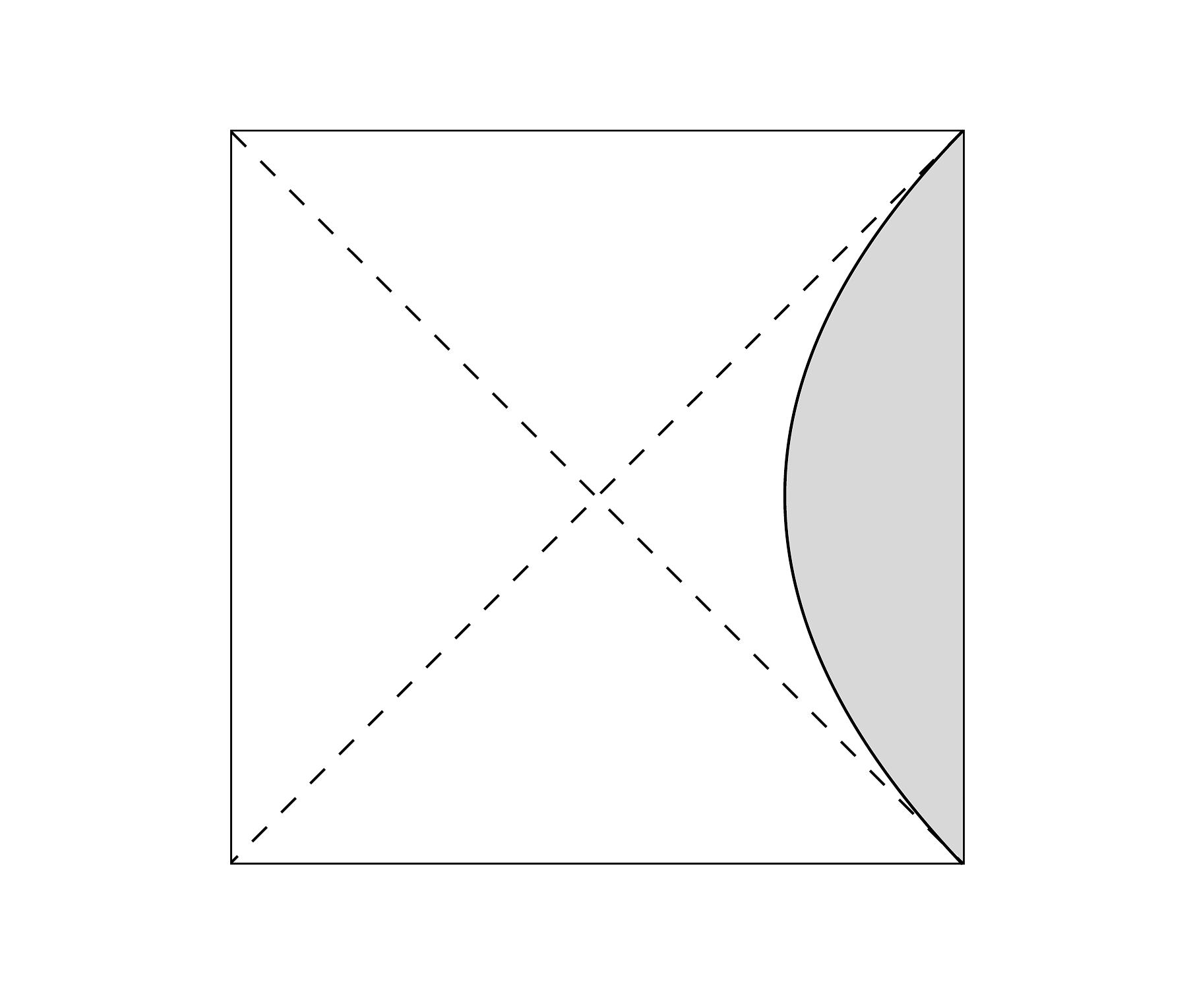
\caption{Bulk region bounded by $\Sigma_r$}
\label{timelike}
\end{minipage}
\end{figure}

The Hamiltonian reduction in Fefferman-Graham gauge on the conformal boundary $\Sigma_\infty$ is well known to lead to Liouville theory \cite{Cacciatori:2001un}. More precisely, the reduction of the entire bulk has two boundaries, one at the future and one at the past boundary. Here, we generalize this result to a Hamiltonian reduction performed over an arbitrary bulk region. We distinguish a piece of bulk bounded by two spacelike surfaces $\Sigma^+_r$ and $\Sigma^-_r$ in the upper and lower diamond, and a piece of bulk bounded by one timelike surface $\Sigma_r$ in either the northern or southern patch, see figures \ref{spacelike} and \ref{timelike}. 

We will carefully derive all steps in the reduction procedure in the upcoming sections. In Section \ref{spec} we will emphasize the new features arising from the reality conditions occurring in the Eddington-Finkelstein gauge instead of the Fefferman-Graham gauge. We will see that it is then convenient to perform a Gauss decomposition of the $SL(2,\mathbb C)$ element far from the identity.

\subsection{Reduction to the non-chiral $SL(2,\mathbb C)$ WZW model}

The first set of boundary conditions allows us to reduce the two Chern-Simons theories to the non-chiral $SL(2,\mathbb C)$ WZW model. Let us start by specifying the boundary terms in the action. We denote the coordinates\footnote{The following derivation does not assume any choice of gauge. The $t$ coordinate might as well be denoted as $u$  in Eddington-Finkelstein coordinates.} as $(t,\phi,r)$, $\phi \sim \phi +2 \pi$ and define $t^\pm = t \pm i \ell \phi$. We define
\bea
S_k[A,\bar A] &=& \frac{k}{4\pi}\int_{Bulk}\, {\rm Tr}\left(A\wedge dA+\frac{2}{3}A\wedge A\wedge A\right) - \frac{k}{4\pi} \int_{\partial Bulk}  dt \, d\phi \, {\rm Tr}(A_t A_\phi),\nn\\
&=& \frac{k}{4\pi}\int_{Bulk} d^3 x\, {\rm Tr}\ \big( 2 A_t F_{\phi r} - \p_t A_r A_\phi +\p_t A_\phi A_r \big)\, .
\eea
Here $\partial Bulk$ is the boundary of the bulk region under consideration at fixed radius $r$ (with one connected component $\Sigma_r$ or two connected components $\Sigma_r^\pm$). Then, the variation of the full action $S_E[A,\bar A]$ given in \eqref{EHaction} is 
\bea 
\delta S_E[A,\bar A] =   \frac{i k }{2\pi} \int_{\p Bulk } dt \, d\phi\   {\rm Tr} \left( A_t \delta A_\phi  -\bar A_t \delta \bar A_\phi   \right) \,  .
\eea
From the first boundary condition, we deduce that a consistent variational principle is given by 
\bea
S_{total} = S_E[A,\bar A]  - \frac{k}{4\pi \ell} \int_{\p Bulk} dt\,  d\phi\ \text{Tr}\big(A_\phi^2 + \bar A_\phi^2\big)\, .
\eea

When $t$ is a timelike coordinate, we observe that $A_t$ is the Lagrange multiplier for the constraint $F_{r\phi} = 0$. This constraint on the initial data implies that locally $A_r$ and $A_\phi$ are pure gauge locally, $A_i = G^{-1} \partial_i G$, $i = r,\phi$ where $G \in SL(2,\mathbb C)$. It is  convenient to choose $A_t = G^{-1}\partial_t G$ as a gauge condition on the Lagrange multiplier $A_t$. We then have 
\begin{equation}
A=G^{-1}dG, \qquad  \bar A =  \bar G^{-1} d \bar G 
\end{equation}
with also $\bar G \in SL(2,\mathbb C)$. We will assume that the decomposition holds globally (no holonomies). When $r$ is timelike, then $A_r$ is the Lagrange multiplier for the constraint $F_{t\phi} = 0$ which we solve as $A_i = G^{-1}\partial_i G$, $i=t,\phi$. We can then gauge fix $A_r = G^{-1}\partial_r G$ and we arrive again at $A=G^{-1}dG$, and similarly $\bar A=\bar G^{-1}d\bar G$.

  
Using the orientation $\eps^{t r \phi} = 1$, we have after imposing the constraint,
\bea
S_k[A] = -\frac{k}{4\pi} \int_{ Bulk}  \frac{1}{3} \text{Tr} (G^{-1}dG)^3 - \frac{k}{4\pi} \int_{\p Bulk} dt \, d\phi\, \text{Tr}\, \left(g^{-1}\p_t g g^{-1} \p_\phi g\right). 
\eea
where we defined $g = (G K)|_{\Sigma_r}$ as the pull-back of $G$ times $K$ defined in \eqref{defK} on $\Sigma_r$. We also define $\bar g = (\bar G \bar K )|_{\Sigma_r}$. 

Therefore, the action is the sum of two chiral WZW models, 
\bea
S_{total} = \frac{ki}{4\pi} S_{WZW}[g] - \frac{ki}{4\pi} \bar S_{WZW}[\bar g] 
\eea
with 
\bea
S_{WZW}[g] &=&\frac{1}{3} \int_{Bulk}  \text{Tr}\, \left(G^{-1}dG\right)^3 + 2 \int_{\p Bulk} dt \, d\phi\ \text{Tr} \, \left(g^{-1}\p_- g g^{-1}\p_\phi g\right),\nonumber\\
\bar S_{WZW}[\bar g] &=&\frac{1}{3}\int_{Bulk}   \text{Tr} \, \left(\bar G^{-1}d \bar G\right)^3 + 2 \int_{\p Bulk} dt\,  d\phi\  \text{Tr}\,  \left(\bar g^{-1}\p_+ \bar g \bar g^{-1}\p_\phi \bar g\right).
\eea
These first order actions describe respectively a right-moving group element $g(t^+)$ and a left-moving group element $g(t^-)$. One thus has $A_- = \bar A_+ = 0$ on-shell. 
The first set of boundary conditions is therefore compatible with the equations of motion of the WZW action. 

Additionally, we could reformulate the combination of two chiral WZW models as one non-chiral WZW model.  To perform this rewriting, one defines $h \equiv g^{-1}\bar g$ and $H \equiv  G^{-1} \bar G = K h \bar K^{-1}$. We observe:
\bea
\frac{1}{3} \int    \text{Tr} \left(H^{-1}dH\right)^3  = -\frac{1}{3} \int   \text{Tr} \left(G^{-1}dG\right)^3 + \frac{1}{3}\int   \text{Tr} \left(\bar G^{-1}d \bar G\right)^3 - \int \text{Tr} \left(d\bar g \bar g^{-1} d g g^{-1}\right)\, .
\eea

We are allowed to trade the variables from $g$ and $\bar g$ to $h $ and $\Pi \equiv -\bar g^{-1}\p_\phi g g^{-1}\bar g - \bar g^{-1}\p_\phi \bar g$. The action then reads
\bea
S_{total} = \frac{ik}{4\pi} \left( -\frac{1}{3} \int  \text{Tr} \,\left(H^{-1}dH\right)^3 +\int dt\,  d\phi\,  \text{Tr}\, \left( \frac{i}{2\ell} \Pi^2 + \frac{i}{2\ell} h^{-1} \p_\phi h \ h^{-1} \p_\phi h \right) +\Pi h^{-1} \p_t h \right)\, .
\eea
Eliminating the auxiliary variable $\Pi$ by its equation of motion, one finally gets
\bea
S_{total} = -\frac{k\ell }{2\pi} \int_{\p Bulk} dt\,  d\phi\,  \text{Tr} \, \left(h^{-1}\p_+ h h^{-1} \p_- h \right) - \frac{ik}{12\pi} \int_{Bulk}  \text{Tr}\,  \left(H^{-1}dH\right)^3\label{2der}
\eea
which is the standard non-chiral $SL(2,\mathbb C)$ WZW action for $h$. It agrees with \cite{Cacciatori:2001un}.

One can express the action in local form upon performing a Gauss decomposition of the form
\begin{equation}\label{gauss2}
H = \left(  \begin{array}{cc} 1 & \hat X \\ 0 & 1 \end{array} \right) \left(  \begin{array}{cc} e^{\frac{1}{2} \hat \Phi} & 0 \\ 0 & e^{-\frac{1}{2} \hat \Phi} \end{array} \right) \left(  \begin{array}{cc} 1 & 0 \\ \hat Y & 1 \end{array} \right) \, ,
\end{equation}
where $\hat X, \hat Y, \hat \Phi$ depend not only on $u,\phi$ but also on $r$. We assume that the decomposition holds globally. For subtleties in the presence of global obstructions, see\cite{Balog:1997zz}. The latter Gauss decomposition allows to rewrite the 3-dimensional integral in  \eqref{2der} as 2-dimensional integrals using the relation 
\bea
\frac{1}{3}\text{Tr} ( H^{-1} dH)^3 = d^3x \ \eps^{\alpha\beta\gamma}\, \p_\alpha \, \left( e^{-\hat \Phi} \p_\beta \hat X\,  \p_\gamma \hat Y \right)\, .
\eea
The 2-dimensional integral in \eqref{2der} can be rewritten equivalently by replacing $h$ by $H|_\Sigma$ since all factors of $K,\bar K$ exactly cancel in the trace. We can then combine all terms (keeping only the radial boundary term) and we find
\bea\label{totalhat}
S_{total} = -\frac{kl}{2\pi} \int_{\p Bulk} dt \, d\phi\,  \left( 2 e^{-\hat \Phi}  \p_- \hat X \, \p_+ \hat Y  +\frac{1}{2} \p_- \hat \Phi \, \p_+ \hat \Phi \right) \, ,
\eea
where all fields $\hat X, \hat Y, \hat \Phi$ have  been pull-backed on $\p Bulk$ which is either $\Sigma_r$ or $\Sigma_r^+ \cup \Sigma_r^-$. 

\subsection{Reality condition and Gauss decomposition}
\label{spec}

Even though the Chern-Simons connection is complex, it describes a real metric and spin connection. Therefore, there is  a reality condition on the connection components, whose precise form depends upon the basis of $SL(2,\mathbb C)$ generators used to express the connection in components. Moreover, there is also a reality condition on the $SL(2,\mathbb C)$ elements $K,\bar K$ used to define the reduced gauge connection. It reflects the fact that the submanifold spanned by $(t^+,t^-)$ is a real submanifold. 

In Eddington-Finkelstein coordinates, we encountered the reality condition
\bea
(EF)\qquad A^\dagger = -\hat \sigma \bar A \hat \sigma,\qquad \hat\sigma^2 = \mathbb I,\qquad \hat \sigma^\dagger = \hat \sigma
\eea
together with $\left(\bar K^{-1}\right)^\dagger \hat \sigma K = \hat\sigma = \bar K^{-1} \hat \sigma \left(K^{-1}\right)^\dagger$, see Section \ref{EFslicing}. 

In Fefferman-Graham coordinates , we encountered the different reality condition 
\bea
(FG) \qquad A^\dagger = \sigma \bar A \sigma,\qquad \sigma^2 = - \mathbb I, \qquad \sigma^\dagger = - \sigma
\eea
together with $\left(\bar K^{-1}\right)^\dagger \sigma K = \sigma = \bar K^{-1} \sigma \left(K^{-1}\right)^\dagger$, see Section \ref{FGslicing}. The matrices $\hat\sigma$ and $\sigma$ were defined in \eqref{defsigmahat} and \eqref{defsigma} respectively. They are defined up to an irrelevant overall sign. 

We expect that there might be other reality conditions in other gauges but we will limit our discussion to two cases above. 

In the case (EF), one finds $\bar G^{-1} = \hat \sigma G^\dagger \tau$ where $\tau \in SL(2,\mathbb C)$ and upon choosing $\tau^\dagger = -\tau$, one has $H^\dagger = -\hat  \sigma H \hat \sigma$. This then implies $h^\dagger = -\hat \sigma h \hat \sigma $. In the case (FG), one finds $\bar G^{-1} = \sigma G^\dagger \tau$ where $\tau \in SL(2,\mathbb C)$ and again upon choosing $\tau^\dagger = -\tau$, one has $H^\dagger = - \sigma H \sigma$. This then implies $h^\dagger = - \sigma h \sigma $.

In case (FG), as discussed in \cite{Cacciatori:2001un}, the matrix $h$ takes the form
\bea
h_{(FG)} =   \left( \begin{array}{cc} u& w \\ -\bar w & v  \end{array}\right)
\eea
with $u,v \in \mathbb R$, $w \in \mathbb C$ and $u v +w \bar w = 1$ while in case (EF), the matrix $h$ takes the form
\bea
h_{(EF)} =  \left( \begin{array}{cc} z& i r_1 \\ i r_2 & \bar z \end{array}\right)\label{Hz}
\eea
with $z\in \mathbb C$, $r_1,r_2 \in \mathbb R$ and $\bar z z + r_1 r_2 = 1$. 

We observe that one can relate these $SL(2,\mathbb C)$ elements as
\bea\label{gauss3}
h_{(EF)} = \hat \sigma h_{(FG)} \sigma 
\eea
which reads in components as
\bea
u = \text{Re} z +\frac{1}{2}(r_1-r_2),\quad 
v  =\text{Re} z -\frac{1}{2}(r_1-r_2) ,\quad w= \text{Im} z + \frac{i}{2}(r_1 + r_2) . 
\eea

The group manifold $SL(2,\mathbb R)$ can be completely covered with the help of 4 coordinate patches. It is natural to use the coordinate patch close to the identity in the case (FG), as done in \cite{Cacciatori:2001un}, using the Gauss decomposition 
\bea\label{gauss1}
h_{(FG)} =  \left(  \begin{array}{cc} 1 & X \\ 0 & 1 \end{array} \right) \left(  \begin{array}{cc} e^{\frac{1}{2} \Phi} & 0 \\ 0 & e^{-\frac{1}{2} \Phi} \end{array} \right) \left(  \begin{array}{cc} 1 & 0 \\  Y & 1 \end{array} \right) \, ,
\eea
where $X,Y,$ and $\Phi$ are function of the coordinates on the slice, $t^+,t^-$. Then, the reality conditions imply $Y = -\bar X$ and $\Phi$ to be real. After imposing the second set of boundary conditions as discussed in the next section,  $\Phi$  will turn out to be the real Liouville field. 

In case (EF) it is then convenient to use the relation \eqref{gauss3} with the Gauss decomposition \eqref{gauss1}. It is easy to see that this coordinate patch for $h_{(EF)}$ does not cover the identity. 

On the one hand, in the (FG) case, comparing the Gauss decompositions \eqref{gauss1} and \eqref{gauss2} and evaluating $K =\bar K^{-1}=\exp(-\log r L_0)$ at fixed $r=r_\Sigma$ we obtain
\bea
\hat X =\frac{1}{r_\Sigma} X , \qquad  \hat Y =\frac{1}{r_\Sigma} Y ,\qquad e^{\hat \Phi} = \frac{1}{r^2_\Sigma} e^{\Phi}. 
\eea 
On the other hand, in the (EF) case, comparing the Gauss decompositions \eqref{gauss3}-\eqref{gauss1} and \eqref{gauss2} and using the values of $K = \bar K^{-1}=\exp (-\frac{i}{2\ell} r L_1)$ at fixed $r=r_\Sigma$ we obtain
\bea
\hat X = X +\frac{i r_\Sigma}{2\ell} , \qquad  \hat Y = Y + \frac{i r_\Sigma}{2\ell} ,\qquad e^{\hat \Phi} = e^{\Phi}. 
\eea

In both cases, the action \eqref{totalhat} reduces to
\bea\label{reduced}
S_{total} = -\frac{k\ell}{2\pi} \int_{\p Bulk} dt d\phi \,  \left( 2 e^{-\Phi}  \p_- X \, \p_+ Y  +\frac{1}{2} \p_- \Phi \, \p_+ \Phi \right) 
\eea 
which is the standard action for the WZW theory. All radial dependence in the action has disappeared. The only possible difference between the Fefferman-Graham and Eddington-Finkelstein cases is the definition of the boundary $\p Bulk$. 

\subsection{Further reduction to Liouville theory}

The second set of boundary conditions on the gauge fields further reduces the WZW  model to a Liouville action.

The boundary conditions were written down in  the language of the gauge field components. Let us first rewrite these boundary conditions in terms of the $SL(2,{\mathbb C})$ element $h$. One way to proceed is to consider the left and right moving WZW currents. They are given by
 \begin{align}
j_a=h^{-1}\partial_a h ,
\qquad \bar j_a=-\partial_a h h^{-1}. 
\end{align}
Using the definition of $h = g^{-1}\bar g$ we deduce 
\begin{align}
j_- &=
 -h^{-1} a_- h + \bar a_-  ,\qquad 
\bar j_+ =  a_+ - h \bar a_+ h^{-1} .
\end{align}
Using the first set of boundary conditions $a_-=\bar a_+=0$, we obtain a simple relation between $h$, the WZW currents, and the gauge fields: $j_- = h^{-1}\partial_- h=\bar a_-$, $\bar j_+ = -\partial_+ h h^{-1}= a_+$. 

For the Fefferman-Graham and Eddington-Finkelstein choices of the $SL(2,\mathbb C)$ generators we have 
\begin{align}\label{setone}
&(FG)\qquad & j_-^1 - i\,  j^2_- &= \frac{2i}{\ell}, & \bar j_+^1 + i \, \bar j_+^2 &= -\frac{2i}{\ell},   & j_-^0 &= \bar j_+^0 = 0 ,\nonumber\\
&(EF ) & j_-^1&=\frac{2i}{\ell}\, ,    & \bar j_+^1&= -\frac{2i}{\ell} \, ,     & j_-^2 &=\bar j_+^2 = 0\, \, .
\end{align}
In either case, the first pair of conditions are first class among themselves. The second pair of conditions can be understood as  a gauge condition for the symmetry generated by the first pair, as discussed in \cite{Henneaux:1999ib,Cacciatori:2001un}.

Using the appropriate Gauss decomposition discussed in the last section, one can rewrite those constraints in terms of the $\Phi,X,Y$ coordinates, with $Y = -\bar X$ and $\Phi$ real. 
In both cases,  (EF) or (FG), the first two constraints are exactly
\bea
e^{-\Phi} \p_- X = \frac{i}{\ell}\, , \qquad 
e^{-\Phi}\p_+ Y = \frac{i}{\ell}\ .
\eea
and the second set of constraints, once combined with the first, becomes
\begin{align}
X= \frac{i\ell}{2}\, \partial_+\Phi\, ,\qquad Y= \frac{i\ell}{2}\, \partial_-\Phi\, .
\end{align}

The constraints are independent of the radius $r_\Sigma$ and independent of the choice of (EF) or (FG) slicing. 

Before inserting the constraints we have to make sure that the action obeys the variational principle. This is the case once we add an improvement term to the action \eqref{reduced}:
\begin{equation}\label{improved}
S_{\rm{impr}}=S_{total}+\frac{k\ell}{2\pi}\int_0^{2\pi} d\phi\ \Big(e^{-\Phi} \left( X\partial_+ Y+Y\partial_-X\right)\Big)\Big|_{t_1}^{t_2}\, .
\end{equation}
After inserting the constraints we are left with the Liouville action
\begin{equation}
S_{\rm{impr}}=-\frac{k\ell}{2\pi}\int_{\p Bulk} dt\, d\phi\ \left(\frac{1}{2}\partial_+\Phi\, \partial_-\Phi + \frac{2}{\ell^2}\exp{\Phi}\right)\ .
\end{equation}
Note that the boundary term in \eqref{improved} contributes as $-2 \frac{k \ell}{2\pi} \int_{\p Bulk} dt d\phi \,  \frac{2}{\ell^2} \exp{\Phi}$. 

The final action is therefore the Liouville action evaluated on the boundary of the bulk region, which can be either  two connected components $\Sigma_r^\pm$ or one connected component $\Sigma_r$, see figures \ref{spacelike} and \ref{timelike}. One can write the Liouville action in covariant form upon coupling it to a metric of Euclidean signature. It is bizarre that when one chooses the radial slice $\Sigma_r$ in the static patch, $t$ is a time coordinate in spacetime, while it is still a Euclidean coordinate of the boundary action. 

\section*{Acknowledgements}

We thank Pujian Mao for his early collaboration on this project. We are grateful to Glenn Barnich, St\'ephane Detournay, Gaston Giribet, Marc Henneaux, Blagoje Oblak and Andrew Strominger for useful conversations.
G.C. is a Research Associate of the Fonds de la Recherche Scientifique F.R.S.-FNRS (Belgium) and he acknowledges the current support of the ERC Starting Grant 335146 ``HoloBHC''. L.D.~is a research fellow of the ``Fonds pour la Formation \`a
la Recherche dans l'Industrie et dans l'Agriculture''-FRIA
Belgium and her work is supported in part by IISN-Belgium and by
``Communaut\'e fran\c caise de Belgique - Actions de Recherche
Concert\'ees''. The work of W.S. was partially supported by the ERC Advanced Grant "SyDuGraM", by a Marina Solvay fellowship, by FNRS-Belgium (convention FRFC PDR T.1025.14 and convention IISN 4.4514.08) and by the ``Communaut\'e Fran\c{c}aise de Belgique" through the ARC program.
\appendix

\section{Conventions}
\subsection*{Orientation}
In coordinates $(t,\phi,r)$ we fix the orientation as $\eps^{r t\phi} = -1$. We use Lorentzian signature so $\eps_{r t \phi}=1$. In order to use Stokes' theorem
\bea
\int_{Bulk} d^3 x \, \p_r (\ldots)= \int_{\p Bulk} d^2x \, (\ldots)  \, ,
\eea
we use $\eps_{r t \phi} = \eps_{t \phi}$. Therefore, $\eps_{t \phi} = 1$. 

\subsection*{$SL(2,{\mathbb C})$ basis}
In the main text we use real  $SL(2,{\mathbb C})$ generators:
\bea
L_0 =\frac{1}{2} \left( \begin{array}{cc} 1&0\\0&-1\end{array}\right),\qquad
L_1 = \left( \begin{array}{cc} 0&0\\1&0\end{array}\right),\qquad
L_{-1} = \left( \begin{array}{cc} 0&-1\\0&0 \end{array}\right)
\eea
with the commutation relations given by
\bea
 [L_0,L_1]=-L_1\, ,\qquad [L_0,L_{-1}]=L_{-1}\, ,\qquad [L_1,L_{-1}]=2L_0\, .
\eea
Furthermore, we define the automorphism of the algebra $\hat \sigma$ as 
\bea
\hat \sigma(L_{-1}) = -L_1,\qquad \hat \sigma(L_1) = - L_{-1}, \qquad \hat \sigma(L_0) =- L_0
\eea
where $\hat \sigma( a) = \hat \sigma^{-1} a \hat \sigma$.  This automorphism exchanges the raising and lowering Lie algebra elements. We also refer to the $SL(2,\mathbb C)$ element $\hat \sigma = i (L_1 + L_{-1})$ with the same notation. 

\bibliographystyle{utphys}

\providecommand{\href}[2]{#2}\begingroup\raggedright\endgroup

\end{document}